\documentclass[journal=langd5,manuscript=article]{achemso}
\usepackage[version=3]{mhchem} 
\usepackage{mathtools}
\usepackage{color}
\usepackage{lmodern}

\author{Kai Huang}
\affiliation
{Materials Science Program, University of Wisconsin-Madison, Madison, Wisconsin 53706, USA}
\author{Sebastian Gast}
\affiliation
{Institute of Chemical Engineering, University of Stuttgart, Stuttgart 70199, Germany}
\author{C. Derek Ma}
\author{Nicholas L. Abbott}
\affiliation
{Department of Chemical \& Biological Engineering, University of Wisconsin-Madison, Madison, Wisconsin 53706, USA }
\author{Izabela Szlufarska}
\email{szlufarska@wisc.edu}
\affiliation
{Department of Materials Science \& Engineering, University of Wisconsin-Madison, Madison, Wisconsin 53706, USA}
\title[\texttt{achemso} demonstration]
{Comparison between Free and Immobilized Ion Effects on Hydrophobic Interactions: A Molecular Dynamics Study}
\begin{document}
\singlespacing
\begin{abstract}

Fundamental studies of the effect of specific ions on hydrophobic interactions are driven by the need to understand phenomena such as hydrophobically driven self-assembly or protein folding. Using $\beta$-peptide-inspired nano-rods, we investigate the effects of both free ions (dissolved salts) and proximally immobilized ions on hydrophobic interactions. We find that the free ion effect is correlated with the water density fluctuation near a non-polar molecular surface, showing that such fluctuation can be an indicator of hydrophobic interactions in the case of solution additives. In the case of immobilized ion, our results demonstrate that hydrophobic interactions can be switched on and off by choosing different spatial arrangements of proximal ions on a nano-rod. For globally amphiphilic nano-rods, we find that the magnitude of the interaction can be further tuned using proximal ions with varying ionic sizes. In general, univalent proximal anions are found to weaken hydrophobic interactions. This is in contrast to the effect of free ions, which according to our simulations strengthen hydrophobic interactions. In addition, immobilized anions of increasing ionic size do not follow the same ordering (Hofmeister-like ranking) as free ions when it comes to their impact on hydrophobic interactions. The immobilized ion effect is not simply correlated with the water density fluctuation near the non-polar side of the amphiphilic nano-rod. We propose a molecular picture that explains the contrasting effects of immobilized versus free ions.

\end{abstract}

\section{Introduction}

Hydrophobic interactions have been recognized as a key driving force for water-mediated self-assembly
processes~\cite{ss1,ss2,ss3,ss4,ss6,ss8} such as protein folding and micelle formation. Because of the ubiquitous nature of ions in biological environments, much research has been dedicated to understanding the effects of specific ions on hydrophobic interactions~\cite{beyond}. While previous studies were mostly focused on the effects of soluble salts, a less explored effect (although of similar importance) is that of proximally immobilized ions. Immobilized charged or polar residues are often present on surfaces of macromolecules, where they are distributed within or are adjacent to nonpolar domains. The impact of these residues on hydrophobicity of the neighboring domains is key to understanding hydrophobic interactions in complex biological environments. Recent measurements based on the atomic force microscopy (AFM)~\cite{Acevedo,Derek} have revealed that the strength of hydrophobic interactions can be modulated by the presence of proximally immobilized ions. This effect was found to be sensitive to the three dimensional nano-patterning of the charged and non-polar groups~\cite{Pomerantz4} and the specific charge (ion) type. Interestingly, the measurable effects of proximally immobilized ions were interpreted to extend over one nanometer in distance, i.e., they are long-range. This finding is in contrast to the recently reported~\cite{Funkner, Stirnemann} short-range nature (i.e., limited to the first ionic hydration shell) of the specific ion effects of soluble salts. The above AFM experiments raise a number of new questions related to specific ions effects. Do the specific proximally immobilized ions follow the same ranking as a function of ion size (Hofmeister order)~\cite{Hofmeister} as free ions? For the same type of ion, how does its influence change when it is transformed from a free ion to a proximally immobilized ion? Without the freedom for ion segregation or depletion from the non-polar domain, what is the molecular origin of the effects of proximally immobilized ions on hydrophobic interactions?

As a first step towards providing insights into these questions, we report MD simulations of hydrophobic interactions between a non-polar surface and a non-polar or amphiphilic molecule in the presence of proximal charges or free ions. Inspired by the experimental studies that use oligo $\beta$-peptide~\cite{beta1,beta2,beta3} that exhibits a well-defined helical conformation, we perform our simulations with a model nano-rod that has a number of key features in common with the experimental system. Specifically, our nano-rod has a well-defined shape and side groups that can be arranged to mimic globally amphiphilic (GA) and non-gobally amphiphilic ({\it iso}-GA) oligo-$\beta$-peptides. In addition, we construct a reference hydrophobic (HP) nano-rod with all side groups being non-polar. There is no corresponding purely hydrophobic $\beta$-peptide studied in experiment as it would be impractical to purify such molecules due to their physical properties. Nevertheless, such HP nano-rods in simulation serve as a useful reference system, allowing us to evaluate the effects of proximally immobilized ions and soluble salts. 

In the remainder of this paper, we present potential of mean force (PMF) calculations of the model nano-rods near a non-polar plate. We first explore the interaction between an HP nano-rod and the extended non-polar plate as a reference system. We discuss the structure of the PMFs and thermodynamics of the hydrophobic interaction. We then investigate the effect of free ions (dissolved salts) by adding alkali halide salts to modify the hydrophobic interaction involving the HP nano-rod. We analyze the water structure and dynamics near the nano-rod, in search for a potential descriptor of the specific free ion effect. We revisit the anomalous effect of lithium and use controlled simulations to investigate different hypotheses for molecular origin of this anomaly. To study the effect of immobilized ions, we replace some of the non-polar side groups of the HP nano-rod by ionic groups. We construct both $iso$-GA and GA nano-rods to explore the effect of surface nano-patterns. For the GA nano-rods, we further study the specific effects of immobilized ions which are long-ranged in nature. Finally, we compare the immobilized and the free ion effects and discuss the origins of their differences.

\section{Molecular Model and Simulation Methodology}
All MD simulations are performed using the GROMACS software package~\cite{gromacs} with explicit solvent since the details of water structuring are key to accurate modeling of hydrophobic interactions. We use the SPC/E force field~\cite{spce} to model water. Long-range electrostatic interactions are calculated using the particle mesh Ewald (PME) method. All hydrogen bonds are constrained through the LINCS~\cite{lincs} algorithm to enable a simulation time step of 2 fs.
The non-polar surface is modeled by 31 Lennard-Jones (LJ) particles arranged into a flat plate. The particles are arranged in a triangular lattice with a lattice constant of 0.32 nm. The same LJ particles (except for their different arrangement) are used to represent the non-polar groups of the nano-rod, whereas the ionic groups are modeled as monoatomic ions. Each of the nine side groups of the nano-rod is bonded to one of the three backbone residues of the molecule and the side group-backbone-side group angles are 120 degree. The backbone residues do not interact with water since they are buried inside the peptide, but their presence allows us to arrange the functional groups in a controlled manner, resembling the rigid helical structure of $\beta$-peptide~\cite{QC}. 

\begin{figure}
\begin{tabular}{cc}
\includegraphics[scale=0.36]{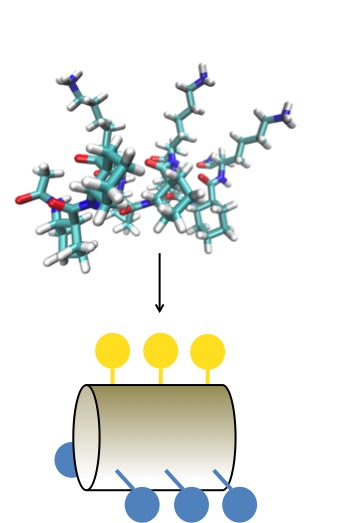} &
\includegraphics[scale=0.32]{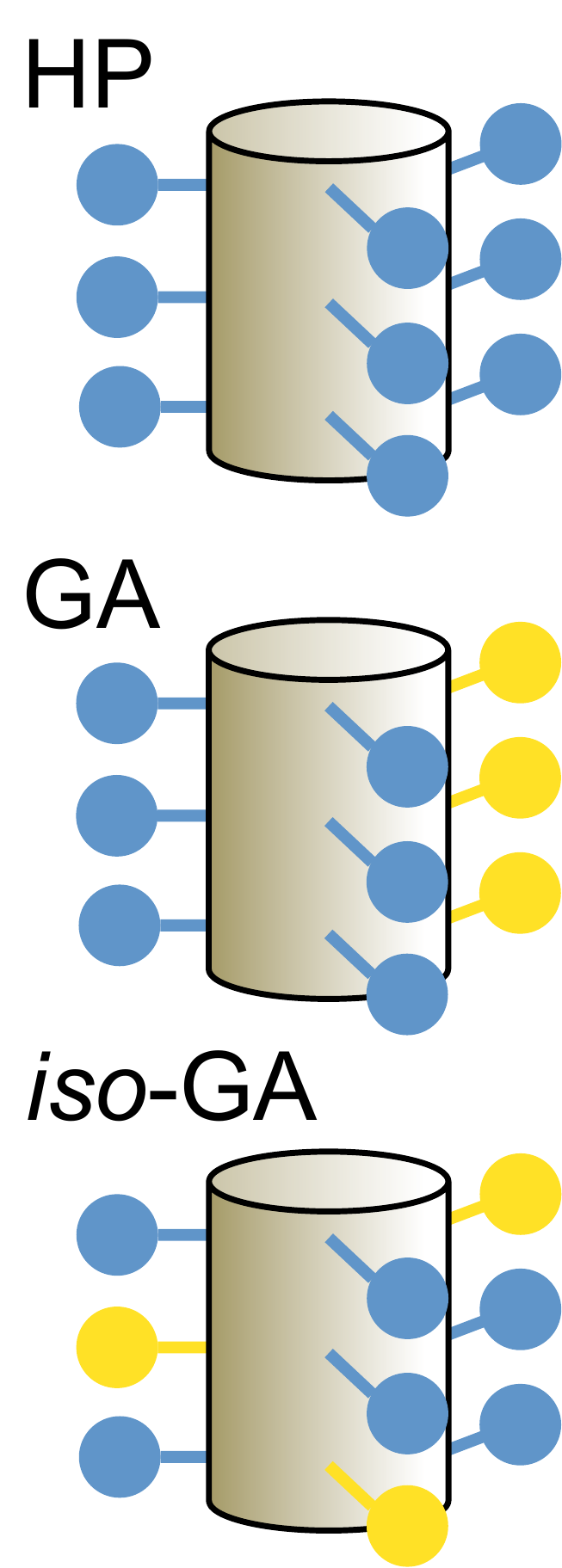} 
\includegraphics[width=0.68\textwidth,viewport=-100 0 1200 800,clip=]{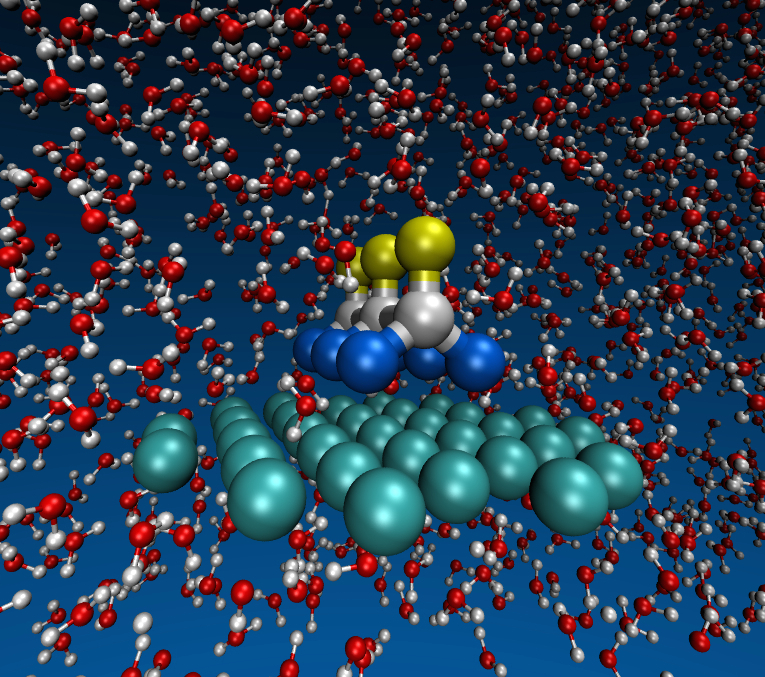}
\end{tabular}
\caption{Coarse-graining of a $\beta$-peptide into a nano-rod with three surface nano-patterns: hydrophobic (HP), globally amphiphilic (GA) and non-globally amphiphilic ({\it iso}-GA). The nano-rod is immersed in water and interacts with an extended non-polar surface. Coarse-grained sites of the non-polar surface are shown in cyan. Non-polar side groups of the nano-rod are shown in blue, backbone in white, and the immobilized ions in yellow. The backbone refers to the coarse-grained residues to which the functional groups are attached.}
\label{system}
\end{figure}

As shown in Fig.~\ref{system}, we construct nano-rods with three different nano-patterns: HP, GA and {\it iso}-GA. Each nano-rod resembles a triangular prism that is approximately 1 nm long. Each edge of the triangular side has a length of approximately 0.5 nm. Similar coarse-grained models have been used in earlier simulations of the self-assembly of $\beta$-peptides~\cite{sbeta2}, but in that case implicit solvent was used. To model the proximal ion effects, we use hypothetical halide ions as ionic side groups tethered to the coarse-grained backbone of the nano-rod. These immobilized halide ions have the same charge and size as the free halide ions to allow a better comparison between immobilized and free ion effects. We model both immobilized and free ions with the OPLS force field (parameters of this force field can be found in Ref ~\cite{opls}). Lorentz-Berthelot mixing rules are applied to generate LJ parameters between different types of atoms. Other LJ parameters used in our study are listed in Table~\ref{t.lj}. The cut-off distance for the LJ interaction and for short-range Coulomb interaction is 1.2 nm. The strength of the hydrophobic interaction between the nano-rod and the extended surface is largely determined by the parameter $\epsilon$ of the LJ potential for interactions between the non-polar site (NS) of the nano-rod and the oxygen atom of water, as well as between the extended nonpolar surface (ES) and the oxygen atom of water. We chose $\epsilon$ for NS-O and ES-O to be 0.6 kJ/mol, which represents a typical hydrophobicity of hydrocarbon molecules~\cite{zangi}. The nano-rod and non-polar surface are solvated in a 2.5 nm$\times$2.5 nm$\times$6 nm water box (around 1100 water molecules). By carrying out two sets of simulations (with three counter-ions added or absent from the solutions) we found that such counter-ions have a negligible effect on the calculated free energy, indicating that the effect of highly dilute counter-ions and their interaction with immobilized ions can be reasonably neglected (see the Supporting Information). When investigating free ion effects, 1 molar concentration (approximately 80 ions) of various alkali halides is added to the solution. Such concentration has been used in previous simulation work to study the free ion effect and in past simulations the Hofmeister ordering of free ions was found to be independent of the concentration of ions for neutral non-polar hydrophobes~\cite{Thomas, Schwierz}.
\begin{table}
\caption{Parameters of the Lennard-Jones potential. H: hydrogen atom of water, O: oxygen atom of water, BB: coarse-grained backbone atom of nano-rod, NS: coarse-grained non-polar site of nano-rod, PC: coarse-grained immobilized ionic group of nano-rod, ES: coarse-grained hydrophobic atom of the extended surface. * symbol refers to atom types that are different from the ones before the hyphen symbol}
\label{tbl:example}
\begin{tabular}{lll}
\hline
Atom types & $\epsilon$ (kJ/mol) & $\sigma$(\AA)\\
\hline
H-* & 0.0 & 0.0\\ 
BB-* & 0.0 & 0.0\\
NS-O & 0.6 & 3.52\\
ES-O & 0.6 & 3.52\\
NS-ES & 0.1 & 4.0\\
PC-ES & 0.1 & 4.0\\
\hline
\end{tabular}
\label{t.lj}
\end{table}

We use the umbrella sampling method to calculate the PMF between each nano-rod and non-polar surface. We define the reaction coordinate to be the $z$ projection of the distance between the second backbone atom (in the middle of the nano-rod) and the non-polar surface, where $z$ direction is perpendicular to the surface. The separation between neighboring sampling windows is 0.05 nm and the spring constant of the external harmonic constraint is 4000 kJ/mol$\cdot\mathrm{nm}^2$. The PMF reported in the main text corresponds to nano-rods that can freely rotate. In addition we have carried out controlled simulations where the rotational degree of freedom is forbidden (details in Supporting Information). We found that the rotation of the molecules does not impact the qualitative conclusions reported in this study. We have calculated uncertainties in PMF using the bootstrapping method~\cite{bootstrap}. Other methods for calculating uncertainties in PMF have also been reported in literature~\cite{zhu}. A 20 ns sampling time for each PMF calculation allowed us to reach an uncertainty smaller than 0.1 kcal/mol.
In our simulations, the system is first relaxed for 1~ns at 300~K and 1 atmosphere using constant pressure constant temperature ensemble with coupling constants $\tau=1$~ps for both the thermostat and the barostat. The velocity rescaling thermostat is used for temperature coupling and the Berendsen barostat with the compressibility of $4.5\times10^5~\mathrm{bar}^{-1}$ is used for pressure coupling.

\section{Results and Discussion}

\subsection{Hydrophobic interaction between HP nano-rod and non-polar plate }

\begin{figure}
\begin{center}
\includegraphics[width=0.7\textwidth,viewport=20 0 700 540,clip=]{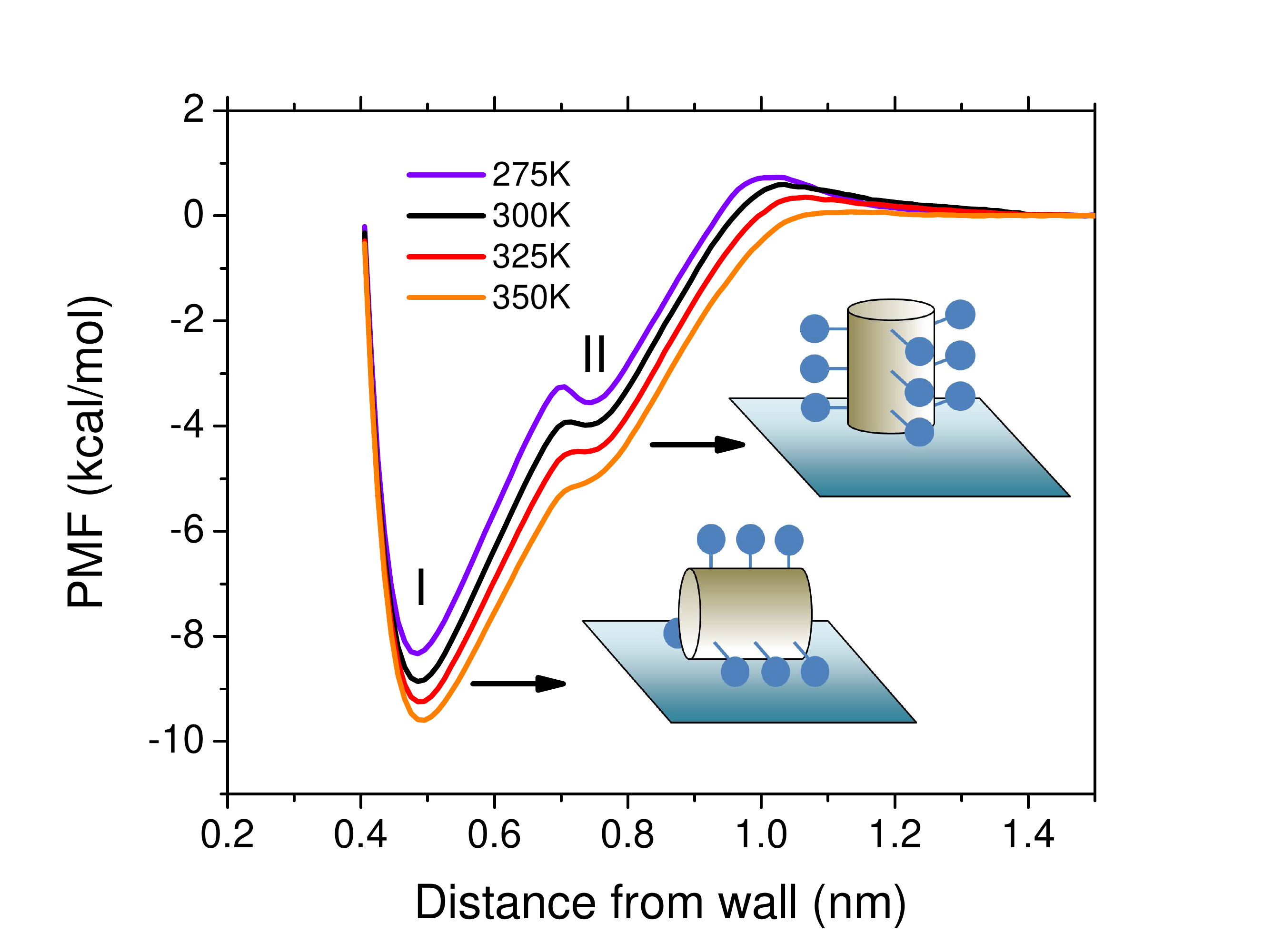}
\caption{Hyrophobic interaction of HP nano-rod as a reference -rods at different temperatures.}
\label{HP}
\end{center}
\end{figure}

Before exploring the specific ion effects, we first study the hydrophobic interaction between a HP nano-rod and an extended non-polar surface as a reference system. Free ion effect can be studied by adding salts to the reference system, whereas proximal charge effect can be investigated by replacing some of the non-polar sites with ionic groups. The PMFs of the hydrophobic interactions for the reference system at different temperatures are plotted in Fig.~\ref{HP}. It is interesting that the overall free energy landscape shown here is more complicated than energy landscapes reported for the interactions between two non-polar solutes of simple shapes~\cite{Thomas, zangi}. While there is only one contact minimum in the attractive part of the PMF in the case of simple solutes, our PMF has two minima (except at temperature higher than 325~K). Specifically, in addition to the primary contact minimum near 0.5~nm, we have a weak secondary minimum near 0.76~nm. The two minima correspond to two energetically favorable configurations of the nano-rod/plate contacts as shown in the insets of Fig.~\ref{HP}. Since there is no electrostatic interaction between the nano-rod and the non-polar surface and the assigned van der Waals interaction is very weak (see Table~\ref{t.lj}), the attractive part of the PMF is largely due to the water-mediated hydrophobic interaction. Therefore, these two contact minima are both hydrophobic in nature. 

It is known that the thermodynamics of the hydrophobic hydration is size dependent~\cite{Chandler1,Blokzijl}. Since the hydrophobic interaction in our system is between a small molecule (nano-rod) and a large surface (non-polar plate), it is interesting to ask what is the thermodynamic driving force of such interaction. According to the partition of the free energy, $\Delta G=\Delta H-T\Delta S$ (where $H$ is enthalpy, $T$ is temperature and $S$ is entropy), the temperature dependence in Fig.~\ref{HP} is indicative of a positive entropy change during the process of association between the nano-rod and the non-polar wall. At $T=300$~K we estimate (based on the free energy difference between $T=275$~K and $T=325$~K) the contribution to the free energy ($-T\Delta S$) to be $-5.4\pm0.9$~kcal/mol, which is more than half of the total free energy change $\Delta G=-8.8$~kcal/mol. The remainder of the free energy change is due to a reduction of enthalpy ($\Delta H=-3.4$~kcal/mol) associated with the interaction. Therefore, the interaction between the HP nano-rod and the non-polar plate is favorable both in entropy and enthalpy. Recall that the association of two macroscopic non-polar surfaces is purely enthalpy driven and association of two small hydrophobes is purely driven by entropy. Our results demonstrate that both of these driving forces (entropy increase and enthalpy reduction) can be present in a mixed system comprised of a small molecule and a large surface. Additional information on the thermodynamic driving force will be presented when discussing Fig.~\ref{iso}. It is worth noting that even for the interaction between small hydrophobes, the thermodynamics can be strongly affected by a nearby flat non-polar surface~\cite{patel2}.


\subsection{Free ion effects}
We now investigate the specific ion effects on the interaction between the nano-rods and the non-polar plate, starting with free ions. The ordering of free ions with respect to their effect on hydrophobic interactions (the so-called Hofmeister series), have been extensively studied in literature~\cite{Thomas, Schwierz} and can be used to test of our simulations. Simulations of free ion effects also allows a direct comparison with the immobilized ion effects that will be discussed later.

Here we investigate the free ion (dissolved salt) effect by carrying out simulations of the interactions between the HP nano-rod and the non-polar plate in the presence of alkali halide salts (NaF, NaCl, NaI, LiCl, CsCl) at modest molar concentration (1M). Before showing the results, we shall however emphasize that the ordering of the Hofmeister series is not trivially determined by the properties of the ions, but also depends on the solute surface~\cite{Schwierz}. For example, the ordering of ions that change the solubility of proteins with a net positive charge can be reversed if the protein becomes negatively charged. As in this paper we are dealing with hydrophobic interaction between charge neutral surfaces, we will use the term Hofmeister series to refer to ordering of ions in the presence of charge neutral hydrophobes. Such Hofmeister series for halide anions with respect to their salting-out ability is $\mathrm{I}^-<\mathrm{Br}^-<\mathrm{Cl}^-<\mathrm{F}^-$~\cite{Schwierz}. For alkali cations, the corresponding ranking is $\mathrm{Cs}^+<\mathrm{Li}^+<\mathrm{K}^+<\mathrm{Na}^+$~\cite{Schwierz}.

\begin{figure}
\begin{tabular}{cc}
\includegraphics[width=0.45\textwidth,viewport=20 0 700 540,clip=]{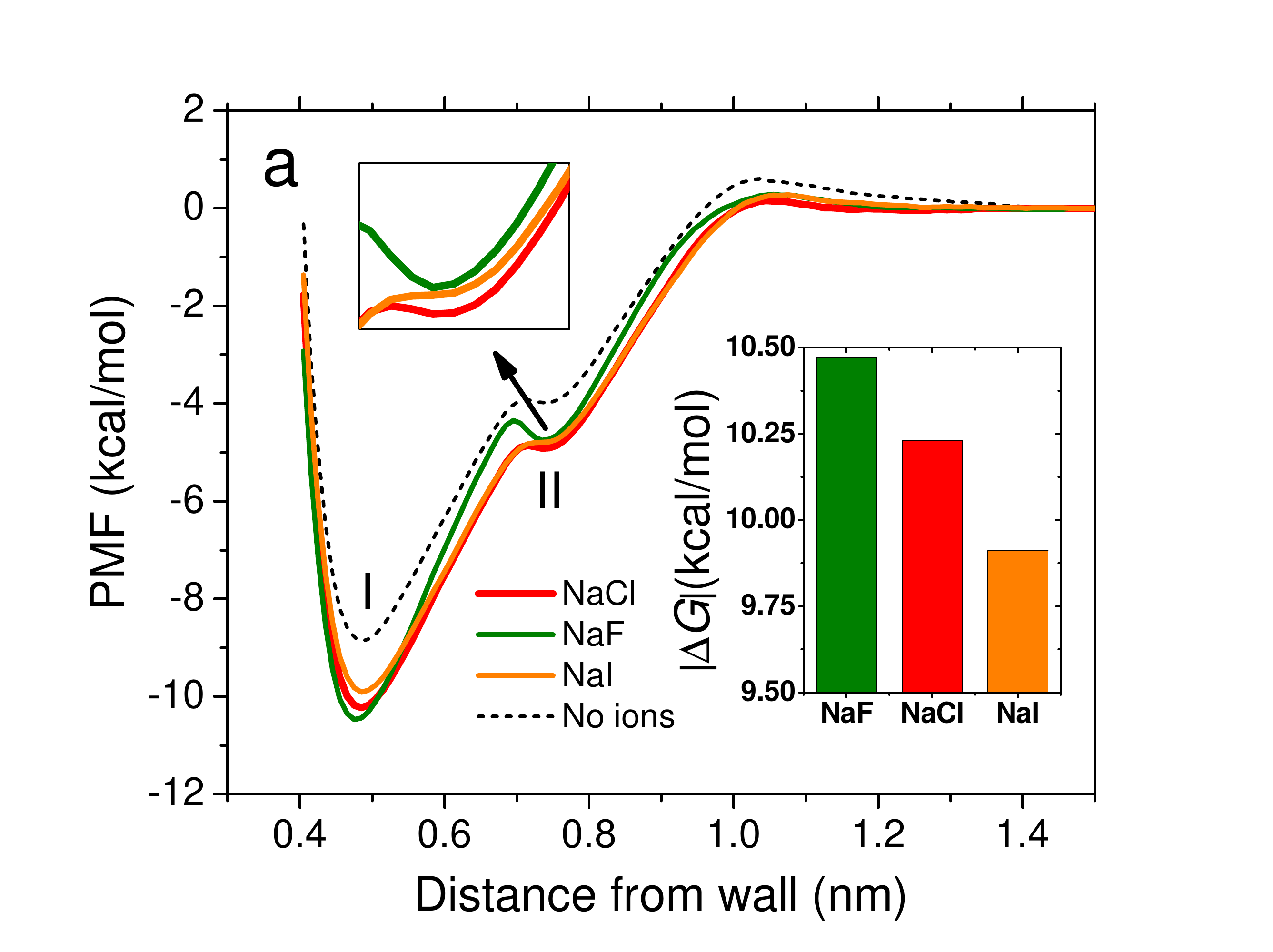} &
\includegraphics[width=0.45\textwidth,viewport=20 0 700 540,clip=]{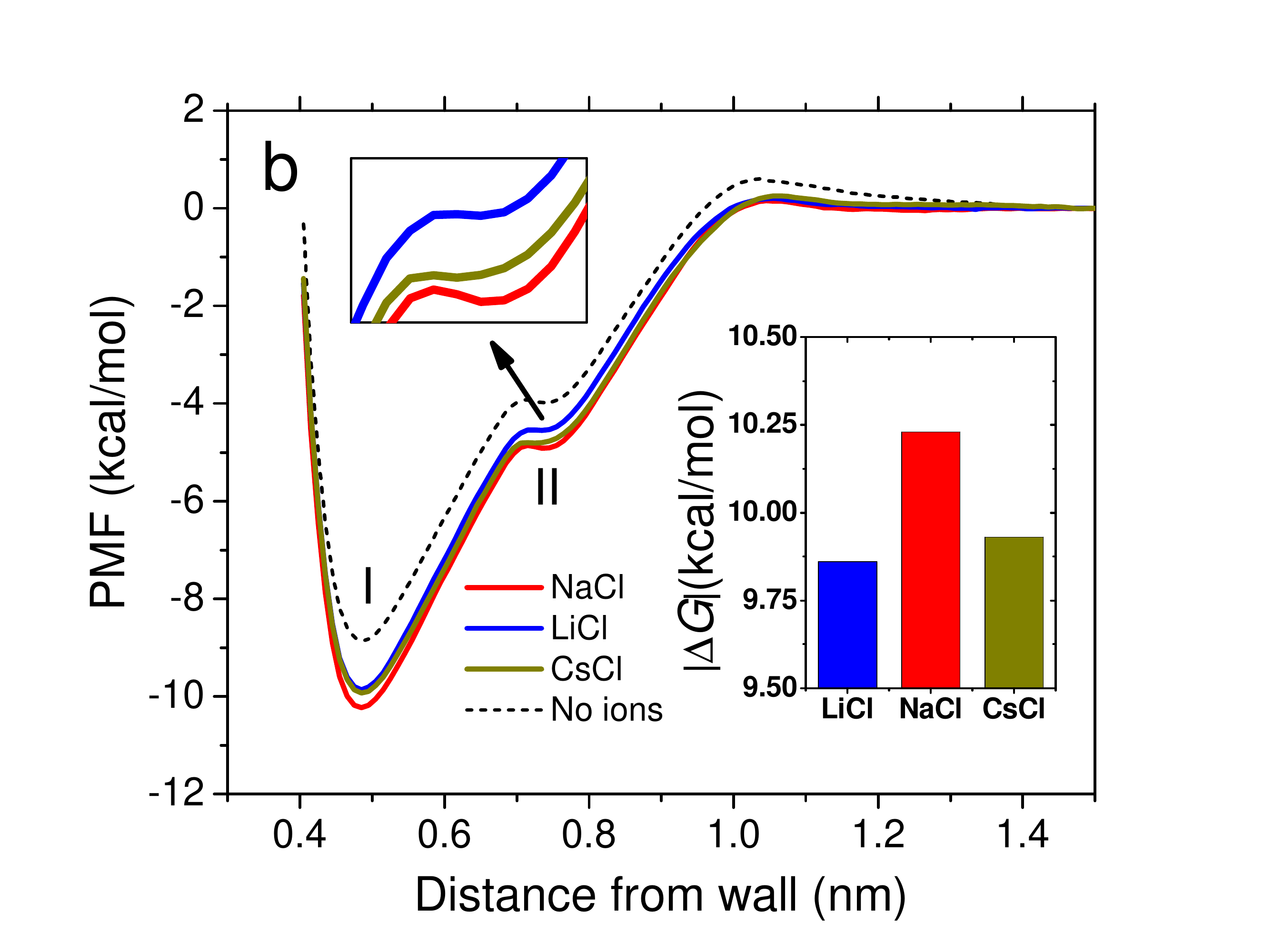}
\end{tabular}
\caption{PMF curves of hydrophobic interaction in the presence of dissolved (a) sodium halide and (b) alkali chloride. The black curve represents the results from a reference system of a HP nano-rod in water in the absence of any free ions or immoblized ions. Primary contact depth plotted in the insets is defined as the absolute value of the difference in PMF between the depth of the primary contact minimum and the reference state where the nano-rod and the surface are separated from each other.}
\label{free}
\end{figure}

Figure~\ref{free} shows the PMFs of the interaction between the nano-rod and the hydrophobic plate in the presence of different salts. As ${\text{Na}^+}$ and ${\text{Cl}^-}$ take positions in the middle of the Hofmeister series and the long-range cooperative ion effects are believed to be small at low and moderate concentrations~\cite{Funkner,Stirnemann}, it is reasonable to assume that the effect of sodium halide (alkali chloride) mainly reflects the specific effect of the halide anions (alkali cations). Comparisons within the sodium halide series and alkali chloride series in Fig.~\ref{free} show that the strength of hydrophobic interaction follows the orders of NaI<NaCl<NaF and LiCl$\approx$CsCl <NaCl. Therefore we can rank halide anions and alkali cations as ${\text{I}^-}$<${\text{Cl}^-}$<${\text{F}^-}$ and ${\text{Li}^+}$$\approx$${\text{Cs}^+}$<${\text{Na}^+}$ in their salting-out effects. Such orderings agree with previous reports of Hofmeister series for non-polar solutes (${\text{I}^-}$<${\text{Br}^-}$<${\text{Cl}^-}$<${\text{F}^-}$, ${\text{Cs}^+}$<${\text{Li}^+}$<${\text{K}^+}$<${\text{Na}^+}$) very well~\cite{Schwierz}. 

\begin{figure} 
\includegraphics[width=0.7\textwidth,viewport=0 0 740 540,clip=]{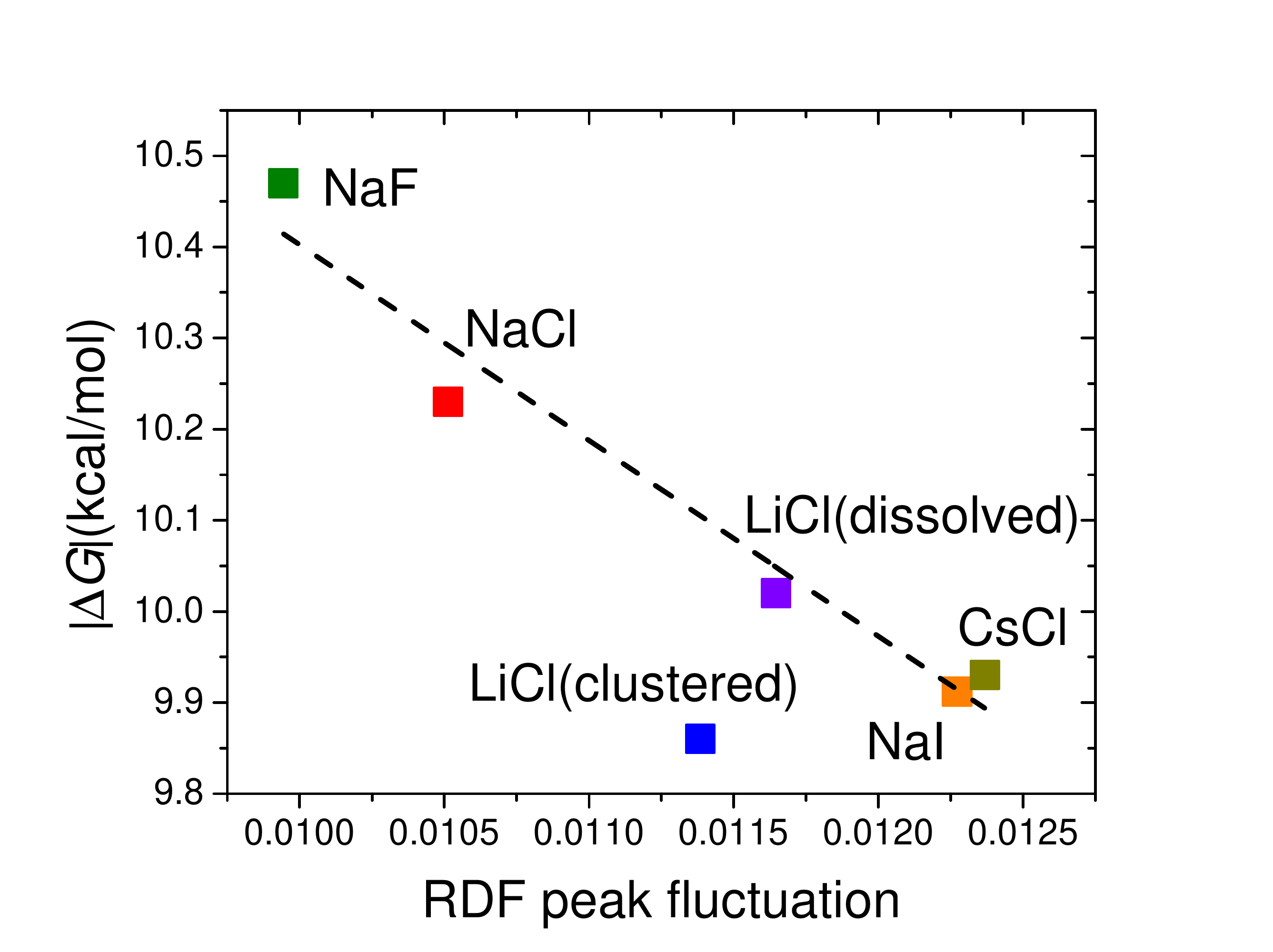}
\caption{Correlation between the strength of hydrophobic interaction in the presence of salts and the RDF fluctuation in the hydration shell of a nano-rod. RDF peak fluctuation is defined as the maximum of the water density fluctuation near the nano-rod.}
\label{linear}
\end{figure}

To shed more light on the molecular origin of the Hofmester ranking of free ions, and inspired by the work of Garde and coworkers~\cite{Jamadagni}, we examine the structure and dynamics of the hydrating water around the HP nano-rod by analyzing the water radial distribution function (RDF) and its time-dependent fluctuation in the presence of different salts (see Supporting Information). Here, the fluctuation of the water density is not merely a measure of the uncertainty of density, but more importantly it is an indicator of the dynamics of water in the hydration shells. For all systems examined in our simulations, we found the RDF of water around the HP nano-rod to be insensitive to the identity of the salts. In contrast, the fluctuation of the RDF is strongly affected by salts. More interestingly, the ion-modulated RDF fluctuation is well correlated with the ion-modulated strength of the hydrophobic interaction ($|\Delta G|$) with an approximate linear relationship as shown in Fig.~\ref{linear}. 

The only significant deviation from this linearity comes from the LiCl salt (see the point labeled LiCl(clustered) in Fig.~\ref{linear}). Anomalous behavior of Li$^+$ has been previously noted in literature. Specifically, while the positions of ions in the Hofmeister series can often be correlated with ion size, lithium ion was found to be an exception to this trend~\cite{Thomas, Schwierz}. Two different ways to rationalize the anomaly of Li$^+$ have been proposed. Thomas and Elcock~\cite{Thomas} postulated that this anomaly is primarily due to clustering between lithium atoms and counter-ions. On the other hand Schwierz {\it et al}~\cite{Schwierz} argued that the anomaly of lithium can be explained by its large effective size if one considers the rigid first hydration shell to be part of the ion. To determine the primary mechanism responsible for the lithium anomaly in our simulations, we carried out an additional calculation of $|\Delta G|$ for LiCl salt solution with a constraint on lithium cations such that Li$^+$ cannot form clusters with chloride anions. The result of this calculation is labeled in Fig.~\ref{linear} as LiCl(dissolved). Interestingly, the salting-out effect (quantified by $|\Delta G|$) of dissociated LiCl is weaker than that of NaCl and NaF and it is stronger than the effect of NaI and CsCl. In other words, even for fully dissolved Li$^+$, its effect on hydrophobic interaction does not correlate with its bare ionic size. Instead, as shown in Fig.~\ref{linear} the effect of dissociated LiC salt follows the same linear trend as other salts, demonstrating that the strength of hydrophobic interactions $|\Delta G|$ is correlated with RDF fluctuations of water around the nano-rod. These results reveal that while the clustering can contribute to the anomaly of lithium, the primary reason for this anomaly is the large effective size of this ion~\cite{Schwierz}. It has been shown in recent literature~\cite{Laage1, Rezus, Tielrooij, Tielrooij2, Funkner, Stirnemann} that large ions (with low charge densities) tend to accelerate the reorientation of water. In this light, our simulation observation of the large RDF fluctuation of water in the presence of Li$^+$ is consistent with the hypothesis that Li$^+$ has a large effective ionic size.

Our conclusion that water fluctuations are a better indicator of hydrophobicity align with the prior results reported by Garde and coworkers~\cite{Jamadagni} for flat surfaces. We demonstrate that these conclusions apply to curved surfaces and more complex molecular structures.

\subsection{Immobilized ion effects} 
Now we turn to the effects of immobilized ions. It has been reported that GA and {\it iso}-GA $\beta$-peptides self-assemble into different structures~\cite{Pomerantz4}, which reflects interactions mediated by the different arrangements of functional groups on the two types of molecules~\cite{Pomerantz1,Pomerantz2,Pomerantz3,sbeta1,sbeta2,sbeta3}. Recent AFM single-molecule force spectroscopy measurements~\cite{Acevedo} revealed that the adhesive forces between $\beta$-peptides and non-polar surfaces depend on the nature of the nano-patterns formed by functional groups on the $\beta$-peptides. Inspired by these experiments, our first goal here is to understand how nano-scale chemical patterns affect intermolecular interactions and how these interactions can be modulated by immobilized ions. To this end, we calculate the PMF between a non-polar plate and nano-rods with different nanoscale patterns. For amphiphilic nano-rods, we use hypothetical $\mathrm{Cl}^-$ ions (see method section) as the proximally immobilized ions in both GA and {\it iso}-GA patterns. 

\begin{figure}
\begin{tabular}{cc}
\includegraphics[width=0.45\textwidth,viewport=20 0 700 600,clip=]{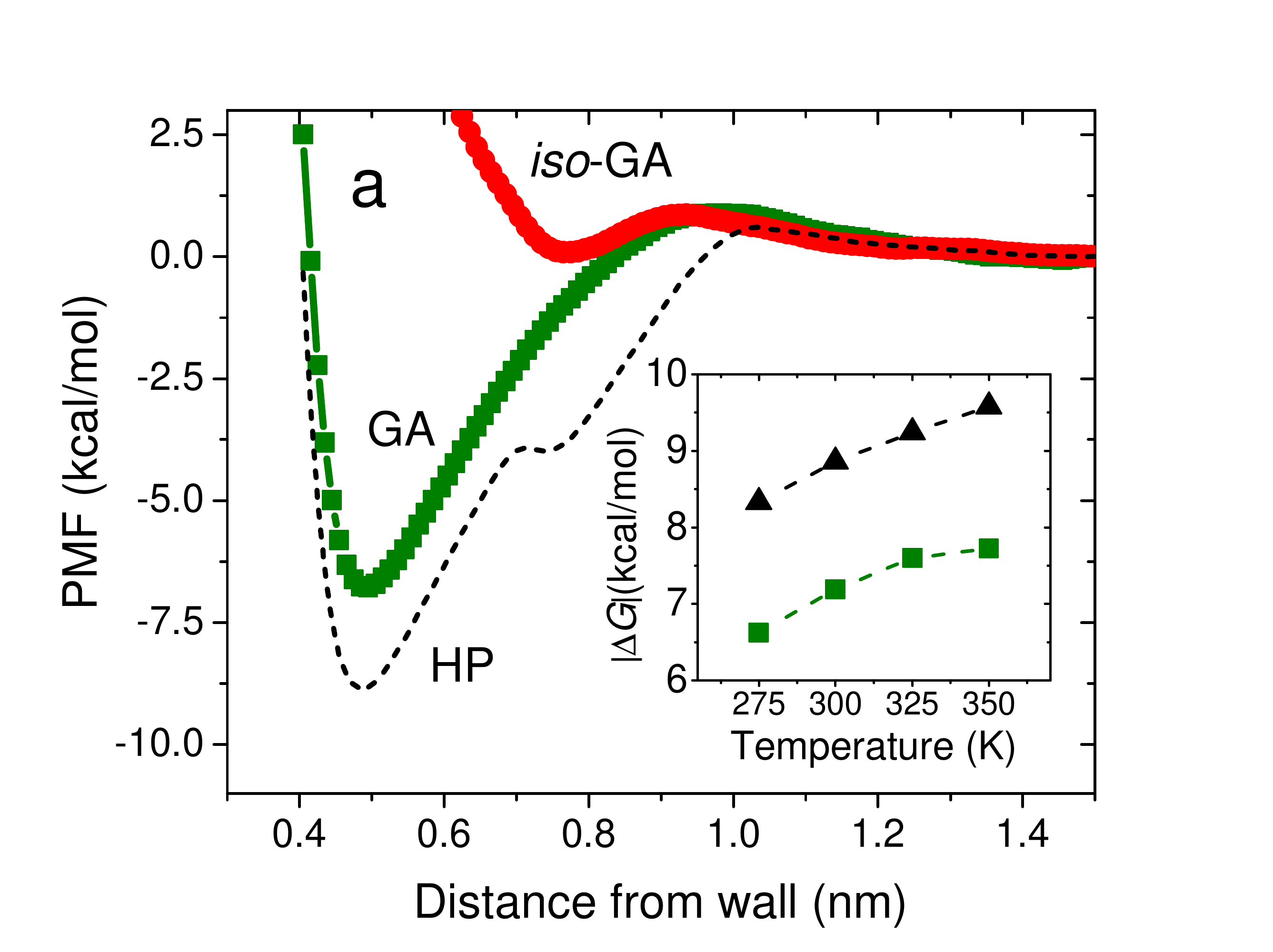} &
\includegraphics[width=0.4\textwidth,viewport=20 0 700 600,clip=]{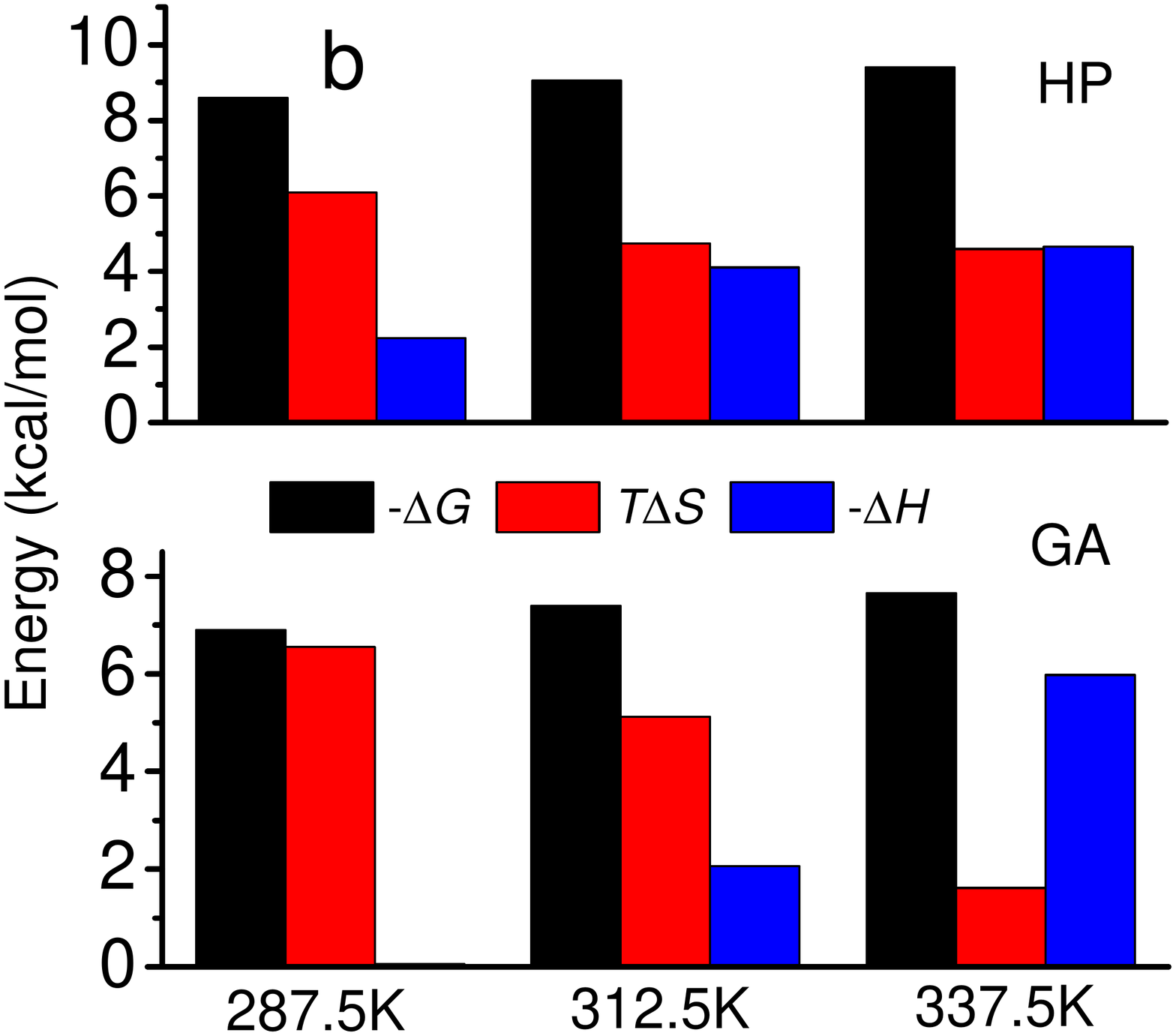}
\end{tabular}
\caption{(a) PMFs of hydrophobic interaction between a non-polar wall and different nano-rods: hydrophobic HP (dashed line), GA (green squares) and {\it iso}-GA (red circles). Inset: The PMF contact minimum at different temperatures: HP (black triangles) and GA (green squares). Contact minimum is defined as the lowest free energy in the PMF with respect to the reference state where that the nano-rod and the surface are completely separated from each other.
(b) The free energy decomposition at different temperatures for the interaction between the non-polar plate and the nano-rods (upper: HP, lower: GA).}
\label{iso}
\end{figure}

The PMF profiles for these two types of nano-rods are shown in Fig.~\ref{iso}a together with a reference PMF (without immobilized ions). All PMF curves are shifted so that the reference state (corresponding to the nano-rod not interacting with the surfaces) has zero free energy. The GA nano-rod interacts with the surface in such a way that six of the nano-rod's non-polar sites face the surface and three of its ionic sites point away from the surface, similarly to the contact type $\text{I}$ of the HP nano-rod shown in Fig. ~\ref{HP}. This arrangement minimizes the exposure of non-polar sites to water and thereby maximizes hydrophobic interaction. For nano-rod with the GA pattern, the primary contact minimum is still pronounced, but it is reduced as compared to the contact minimum for the HP nano-rod. The second minimum, which was observed for the HP nano-rod, disappears in the case of the GA pattern.  In contrast, the PMF of the {\it iso}-GA pattern has a very shallow contact minimum of type $\text{II}$ and is largely repulsive in the regime where the primary minimum (corresponding to configuration $\text{I}$) would occur. Although it is possible for the {\it iso}-GA nano-rod to align itself so that up to four of its non-polar sites face the surface, it is clear that the strength of hydrophobic interaction does not simply scale with the number of non-polar sites that face the surface. The striking difference in the PMF of the GA and the {\it iso}-GA nano-rods reflects the distinct hydration status of the two molecules in water.
More specifically, we find that hydrophobic interaction can be effectively destroyed by adding proximally immobilized ions between non-polar sites. Consequently, our results demonstrate that the hydrophobic interaction is a result of the collective behavior of hydrating water molecules and it is not a simple function of the surface area of non-polar domains. This conclusion is generally consistent with the AFM measurement by Acevedo {\it et al}~\cite{Acevedo}, although in the experiments a non-vanishing pull-off force persisted in the case of {\it iso}-GA $\beta$-peptide. This force was established to be electrostatic in nature due to the non-polar surface accumulating an excess negative charge when immersed in water. The immobilized ion effect in the {\it iso}-GA nano-rod is similar to the one recently reported in MD simulations by Acharya {\it et al}~\cite{Acharya} where it was shown that polar groups in the middle of a flat non-polar domain can substantially modulate hydrophobic interactions. Our results extend Acharya {\it et al}'s conclusion for polar groups to immobilized ions, and from a flat geometry to a nano-rod.

To determine the thermodynamic origin of the hydrophobic interaction predicted by our MD simulations for nano-rods with the GA nano-pattern, we calculate the PMFs for hydrophobic interactions at a number of different temperatures as shown in the inset of Fig.~\ref{iso}a. Based on this temperature dependence, we partition the free energy into the entropy and enthalpy contributions. As shown in the Fig.~\ref{iso}b, around room temperature, both the entropy and enthalpy driving forces are present in the interactions between the non-polar plate and the nano-rods (with and without immobilized ions). In other words, both the HP and GA nano-rods have a negative hydration entropy associated with their non-polar surfaces. The entropy driving force decreases with temperature as the hydrating water becomes more disordered.

Having verified that the hydrophobic interaction involving amphiphilic molecules can be turned on and off by choosing either the GA or {\it iso}-GA nano-pattern of the nano-rod, we now test the possibility of tuning the strength of the hydrophobic interactions by varying the size of the proximally immobilized ions of the GA nano-rod. We also investigate whether the effects of proximally immobilized ion follows the same Hofmeister order as the dissolved free ions, a question that is of practical importance for the rational design of hydrophobically driven self-assembly of materials~\cite{ss1,ss2,ss3,ss4,ss6,ss8}. 

\begin{figure} 
\begin{tabular}{cc}
\includegraphics[width=0.5\textwidth,viewport=0 10 740 540,clip=]{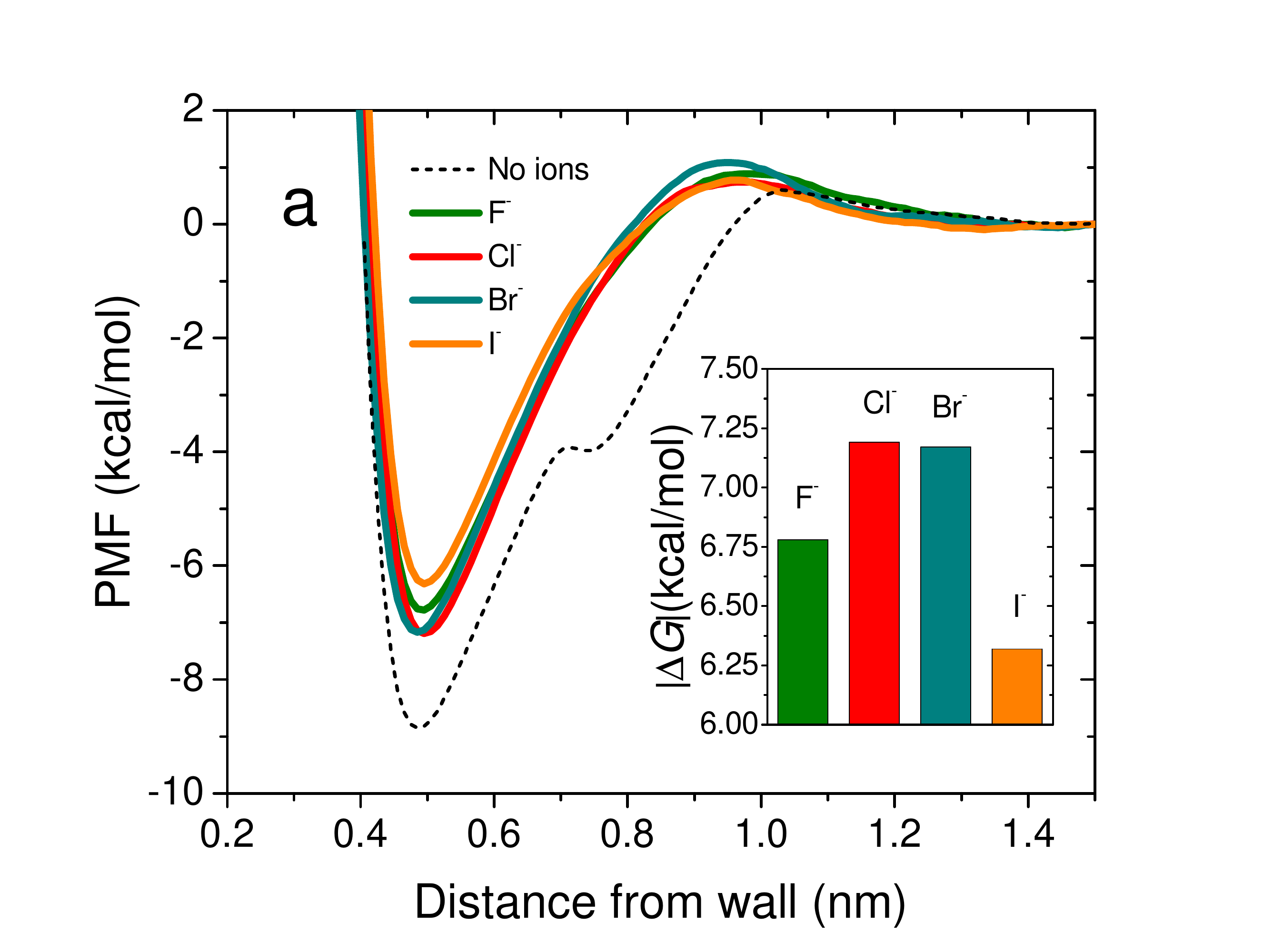} &
\includegraphics[width=0.35\textwidth,viewport=0 -100 500 400,clip=]{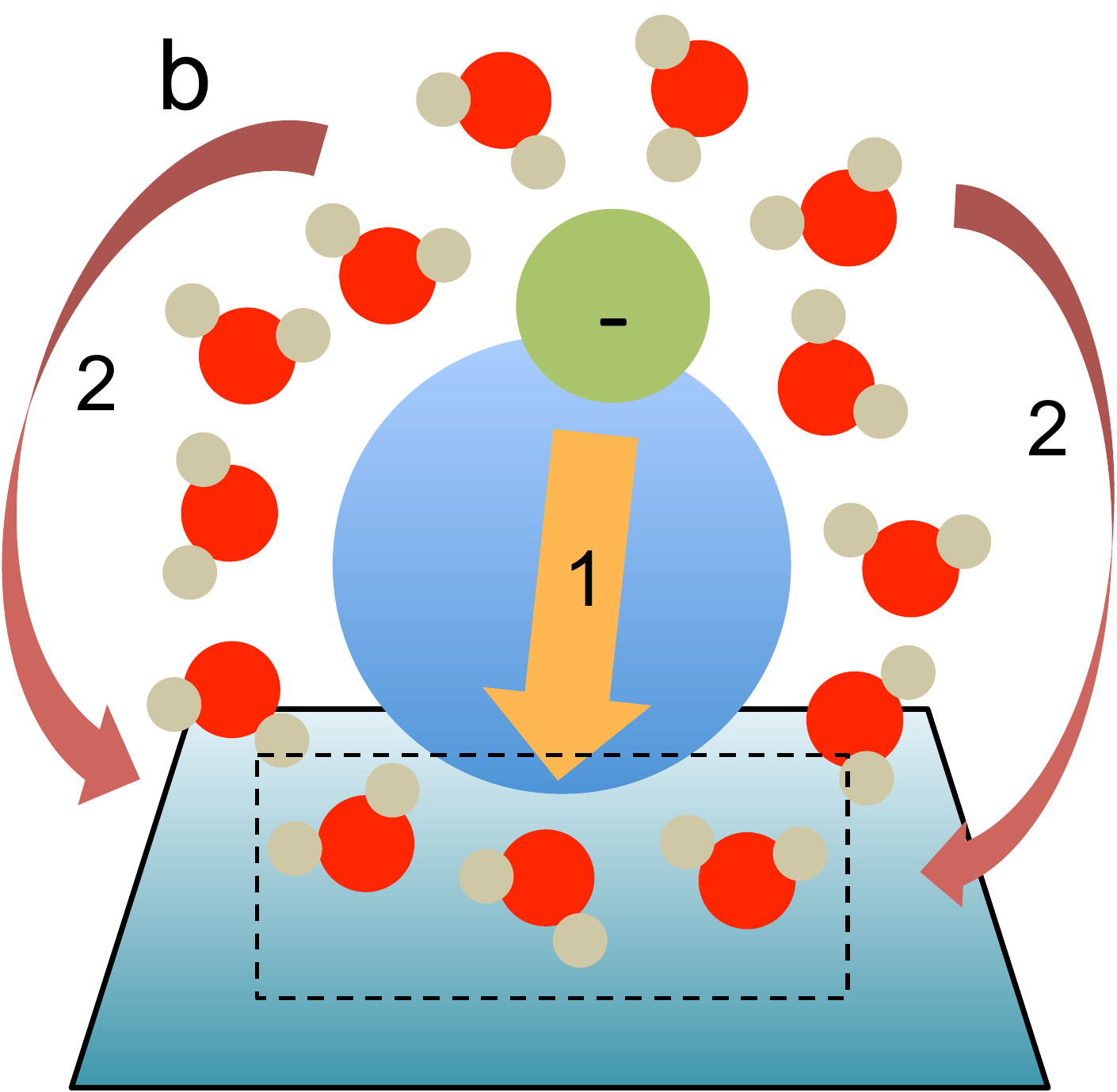}
\end{tabular}
\caption{(a) PMF of hydrophobic interaction between a non-polar wall and the nano-rod with proximally immobilized ions with sizes that are comparable to halogen anions. The charge is constant at -1. The inset shows the depth of the contact minimum, which is defined as the absolute value of the difference in PMF between the contact minimum and the reference state where the peptide and the surface are separated from each other. (b) Schematic molecular picture of specific immobilized ion effect. Straight arrow (1) across the non-polar part of the nano-rod (blue) represents the direct electrostatic (charge-dipole) interaction between the immobilized ions (green) and the interfacial water molecules (in the dashed-line box). Curved arrows (2) represent the indirect long-range perturbation of the structuring of the water molecules by the immobilized ions.}
\label{size}
\end{figure}

To shed light on the question of the effect of the ionic size of a proximally immobilized ion on hydrophobic interaction, we calculate the PMF for the nano-rods with the proximally immobilized ions being halide anions. We choose halide anions for our test because their specific ion effects are known to be stronger than alkali cations, and their Hofmeister order in the case of free ions correlates well with the ionic sizes or, equivalently, charge densities. The PMF results are presented in Fig.~\ref{size}a. Interestingly, we find that the strength of hydrophobic interaction between the nano-rod and the non-polar surface does not depend monotonically on the size of the proximally immobilized ion. Instead, the ranking of the interaction, from weak to strong, with different proximally immobilized ions follows $\mathrm{I}^-<\mathrm{F}^-<\mathrm{Cl}^-\approx\mathrm{Br}^-$. 
To ensure that the ordering of ions is not affected by the possibility of the nano-rod to reorient itself during simulations, we carried out additional simulations where the orientational freedom of the nano-rod is frozen out. In these constrained simulations, all nano-rods have their non-polar surface lying parallel to the plate during the sampling so that the reaction paths are identical for all ions. The results shown in the Supporting Information show the same ordering of ions with the interaction strength, confirming that the rotational freedom of nano-rod (or its absence) does not alter the order of the specific immobilized ion effects. In addition, we have calculated uncertainty of the PMF calculations (as explained in the Section Molecular Model and Simulation Methodology) and it was found to be smaller than 0.1~kcal/mol.

Two types of interactions can hypothetically contribute to the observed effects of immobilized ions on hydrophobic attraction and they are both illustrated by arrows in Fig.~\ref{size}b. One contribution comes from the electrostatic interaction between the immobilized ion and water molecules residing at the nano-rod/non-polar surface interface. This interaction occurs across the diameter of the GA nano-rod. The second contribution comes from long-range perturbation of hydration shells by the immobilized ions. Both of these contributions can affect free energy of the interfacial water (between the nano-rod and the non-polar surface). The electrostatic (charge-dipole) contribution to interaction is sensitive to the charge of the ion and not to its size. On the other hand, the contribution from long-range perturbation of hydration shell can affect the entropy of interfacial water and it is sensitive to the size of the immobilized ion. The overall specific immobilized ion effect is expected to be a balance of these two mechanisms. It is worth pointing out that the water density fluctuation that correlates with the free ion effect is no longer an indicator of the long-range immobilized ion effect. Specifically, analysis (not shown here) of the water density and its fluctuation near the non-polar side of the GA nano-rods shows that these quantities are not simply correlated with the specific immobilized ion effect. The lack of correlation here suggests that perturbation of the free energy of the water beyond the first ionic hydration shell is complicated in nature and further studies are needed to better understand the underlying molecular details.

Although many differences exist between the recent experiment studies using $\beta$-peptides and the models used in our simulations, our simulations do support the experimental finding that the immobilized ions can have a long-range effect on hydrophobic interactions.

\subsection{Comparison between the effects of proximally immobilized ion and soluble salts}

\begin{figure}
\includegraphics[width=0.7\textwidth,viewport=0 0 760 550,clip=]{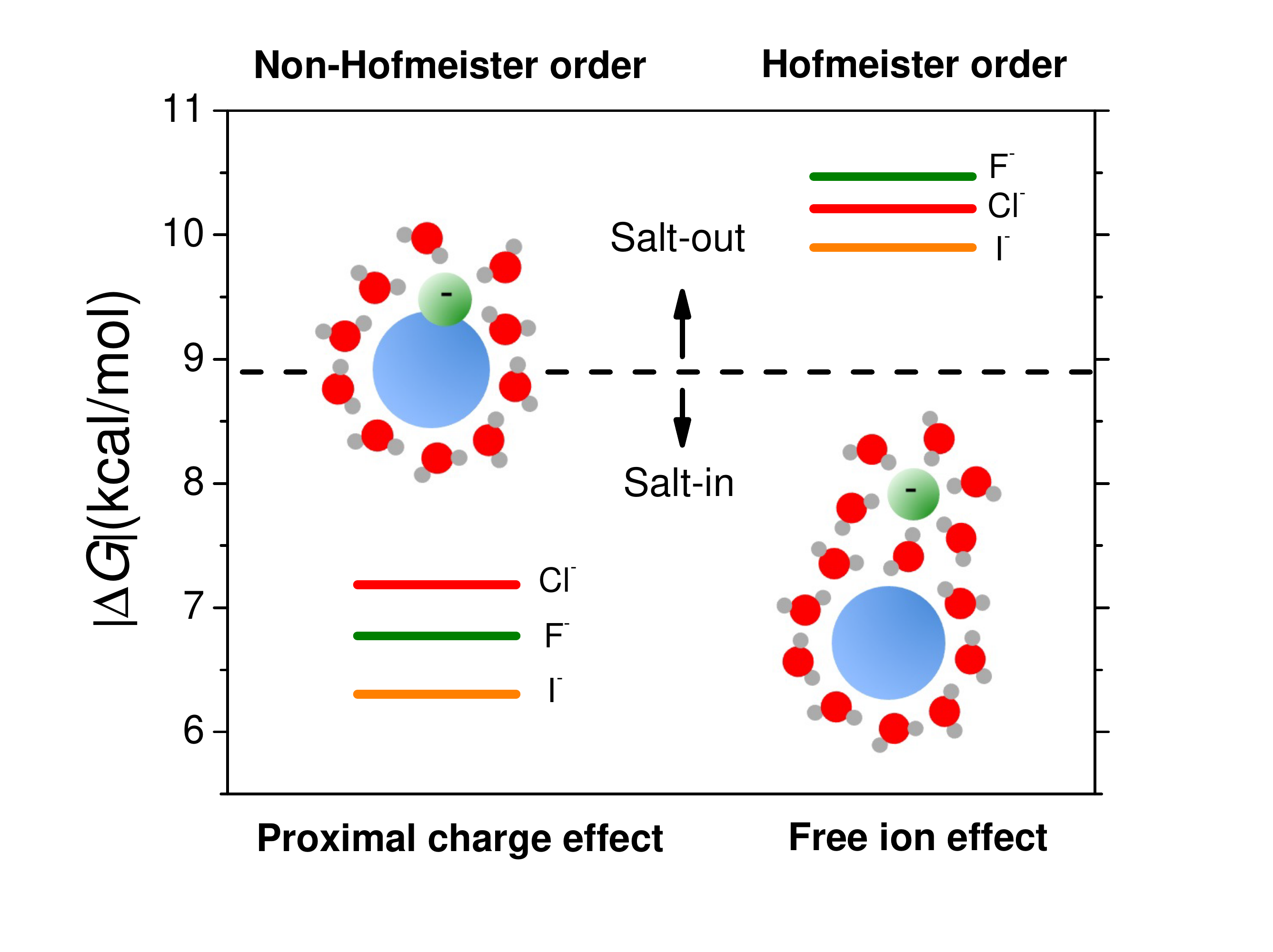}
\caption{Comparison between immobilized and free ion effect.}
\label{compare}
\end{figure}

It is instructive to compare the effects from the same ion in different states, soluble and immobilized. In Fig.~\ref{compare} we summarize the effects of immobilized ions and free ions. We choose the reference (a dashed line in Fig.~\ref{compare}) for both cases to be the hydrophobic interaction strength between the HP nano-rod and the extended non-polar surface without any ions. Inspection of Fig.~\ref{compare} reveals that the immobilized ions tend to weaken the hydrophobic interaction while the free ones tend to strengthen it. This difference is related to the distinct spatial distributions of immobilized and free ions. One can think of the hydrophobic interaction as a dehydrating reaction accompanied by the release of water molecules from the surface of the solute to the bulk region, which can be symbolically written as
\begin{equation}
\text{Nano-rod}\cdot \mathrm{H_2O}+\text{Surface}\cdot \mathrm{H_2O} \xrightleftharpoons[\text{hydrate}] {\text{dehydrate}}\text{Nano-rod|surface}+\mathrm{H_2O}(\text{bulk}),
\label{reaction}
\end{equation}
where symbol $\cdot$ means hydrating and symbol $|$ means in contact.
The strength of hydrophobic interaction is determined by 
\begin{equation}
\begin{split}
\Delta{G}_{\text{dehydrate}}&=G_{\mathrm{H_2O}(\text{bulk})}-[G_{\text{Nano-rod}\cdot\mathrm{H_2O}}+G_{\text{Surface}\cdot\mathrm{H_2O}}-G_{\text{Nano-rod|surface}}]\\
&=G_{\mathrm{H_2O}(\text{bulk})}-G_{\mathrm{H_2O}(\text{surface})}.
\label{G}
\end{split}
\end{equation}
Here, we defined $G_{\text{H2O(surface)}}$, which qualitatively can be understood as the free energy of water near the non-polar surfaces (the non-polar domain of GA nano-rod and the flat non-polar surface) when they are far from each other.
The specific ion effect on hydrophobic interaction can therefore be generally understood as the difference in the manners the ion modifies the free energy of hydrating water and of bulk water
\begin{equation}
\Delta(\Delta{G}_{\text{dehydrate}})=\Delta G_{\mathrm{H_2O}(\text{bulk})}-\Delta G_{\mathrm{H_2O}(\text{surface})}.
\label{ion}
\end{equation}
The sign of $\Delta(\Delta{G}_{\text{dehydrate}})$ determines whether the ion increases or decreases the strength of hydrophobic interaction.
As we are interested in the immobilized ion effect, we here focus on the free energy of water hydrating the nano-rod. The free energy of water hydrating the non-polar plate is not sensitive to the proximally immobilized ions and can be treated as a constant reference because the plate is well isolated from the nano-rod in the non-interacting reference state.
The distinct effects from an immobilized ion and a free ion result from their different spatial distributions with respect to the nano-rod, i.e., immobilized ions are immobilized near the nano-rod and absent in the bulk water whereas free ions are excluded from the nano-rod and remain in the bulk. Therefore in Eq~\ref{ion}, the $\Delta G_{\mathrm{H_2O}(\text{bulk})}$ term is important for free ions but it is negligible for immobilized ions. This means that an immobilized ion weakens the hydrophobic interaction solely by lowering the free energy of water hydrating the nano-rod. In contrast, the free ion effect is a result of the changes in both the hydrating free energy and bulk free energy. In our case, as the nano-rod is small in size and highly convex in curvature, free ions are generally excluded from it (see Supporting Information) and mainly lower the free energy of bulk water. The exclusion of ions, including ${\text{I}^-}$, from small hydrophobes have been recently reported in experiments~\cite{Rankin1, Rankin2}. For bulkier complex ion such as guanidinium, it has been shown that the dispersive interaction between the ion and the non-polar solute can favor the dissolution of the latter~\cite{godawat, breslow}. It is interesting to note that regardless of whether ions are immobilized or dissolved, the same ions have different rankings with respect to their modulations of strength of hydrophobic interactions.

\section{Concluding remarks}

We have used MD simulations to study the effects of proximally immobilized ions on hydrophobic interactions between a $\beta$-peptide inspired nano-rod and an extended non-polar surface. By comparing the effect of immobilized ions with different nano-patterns and ionic radii, we demonstrated that hydrophobic interaction can be largely eliminated by {\it iso}-GA patterning and can be modulated by proximally immobilized ions in the globally amphiphilic (GA) nano-rod. Several trends observed in our simulations agree with recent single-molecule AFM measurements~\cite{Acevedo,Derek} and we provide a molecular mechanistic understanding of these trends in the context of the models used in the MD simulation.
In the broader context of specific-ion effects, we have shown that the immobilized ions do not necessarily follow the same ordering as the free ions. Our analysis of the structure and dynamics of water near the hydrophobic nano-rod shows that dynamics is a better indicator of the specific ion effect than the static water structure. This result extends prior results of Garde and coworkers~\cite{Jamadagni}. Our results provide new insights into specific ion effects that may, in the long term, guide the rational design of hydrophobic interactions and self-assembly process driven by these interactions.

\section{Associated content}
Supporting Information

Figure S1. Radial distribution function (RDF) of free ions around the nano-rod. Figure S2. Radial distribution function (RDF) and its fluctuation for water oxygen around nano-rod. Figure S3. Anomaly of the lithium effect. Figure S4. Comparison between simulations with and without counter-ions. Figure S5. Proximally immobilized ion effect in constrained simulation. This material is available free of charge via the Internet at http://pubs.acs.org.

\begin{acknowledgement}
The authors gratefully acknowledge support from NSF-NSEC at UW-Madison DMR-0832760, from NSF grant CMMI-0747661, and from the ARO (W911NF-14-1-0140).
\end{acknowledgement}
\bibliography{reference}

\newpage
\section{TOC Graphic}

\setcounter{figure}{0} 
\begin{figure}
\includegraphics[scale=0.5]{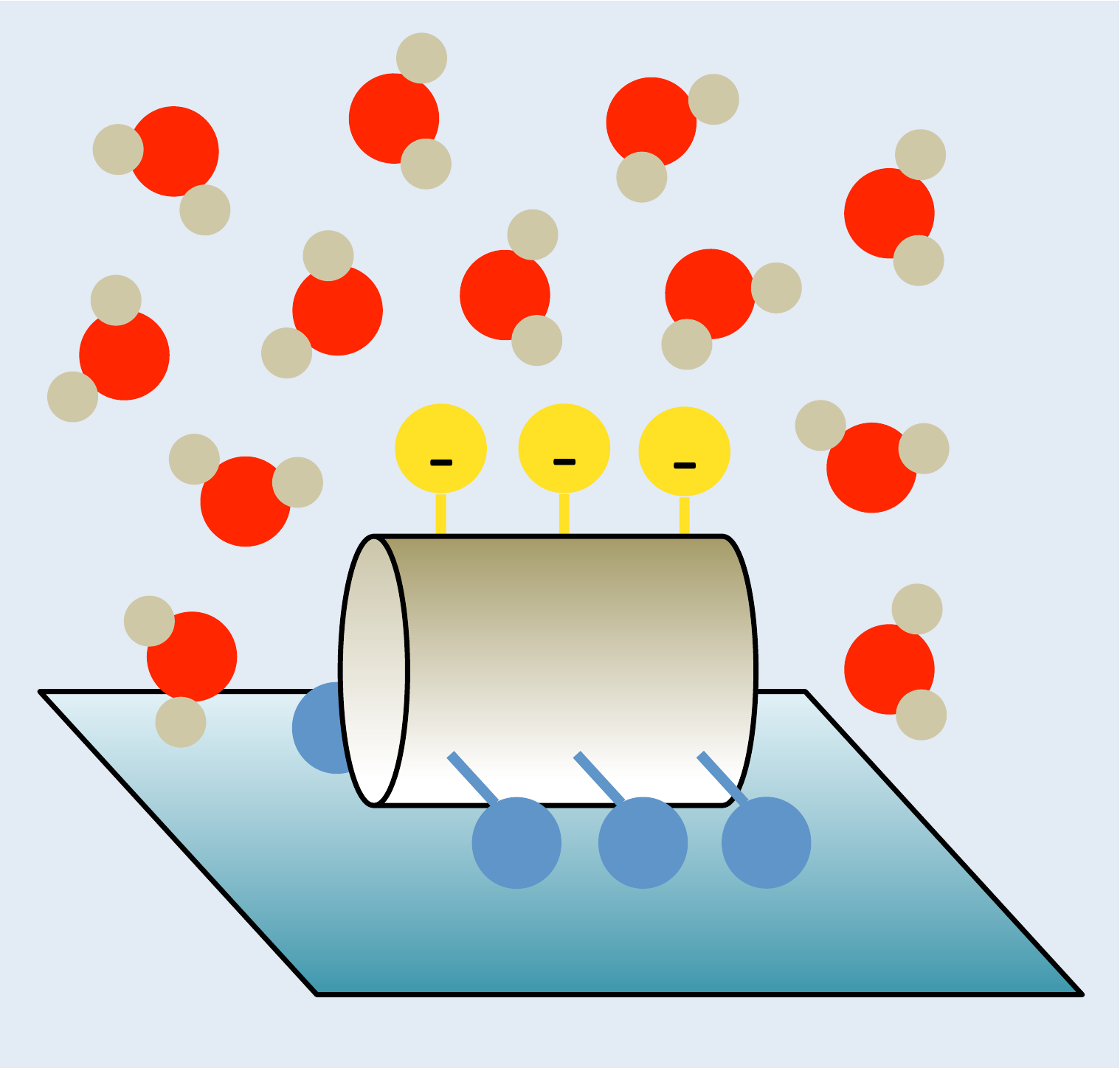}
\label{f.toc}
\end{figure}

\end{document}